\documentstyle[preprint,prd,aps]{revtex}
\begin{document}
\draft

\title{Kramer--Neugebauer Transformation
for Einstein--Maxwell--Dilaton--Axion Theory}

\author{Kechkin Oleg and Yurova Maria }

\address{Nuclear Physics Institute,\\
Moscow State University, \\
Moscow 119899, RUSSIA, \\
e-mail: kechkin@cdfe.npi.msu.su}

\date{\today}

\maketitle

\draft

\begin{abstract}
The Kramer--Neugebauer--like transformation is constructed for the
stationary axisymmetric D=4 Einstein--Maxwell--dilaton--axion system.
This transformation directly maps the dualized sigma--model equations
of the theory into the nondualized ones. Also the new chiral $4 \times 4$
matrix representation of the problem is presented.
\end{abstract}



\draft

\narrowtext

\section{Introduction}
Recently much attention has been given to the study of symmetries admitted
by gravitational models appearing in superstring theory low energy limit
\cite {ss}--\cite {kr}.
As it has been established, one of such models, $D=4$
Einstein--Maxwell--dilaton--axion (EMDA) theory possesses the $Sp(4,R)$
group of transformations in the stationary case \cite {jmp}--\cite {pr1}
and allows the $Sp(4,R)/U(2)$ null--curvature matrix representation
\cite {pr2}--\cite {pl}.

Here we continue to investigate this theory. It is established that
its stationary equations can be written on the language of two symmetric
$2 \times 2$ matrix variables in two different ways.
The first is connected with the employment of the scalar
matrices $P$ and $Q$ \cite {pl}
which are the Gauss decomposition components of the $Sp(4,R)/U(2)$ matrix
$M$. The matrix $Q$ is defined by the set of dualized Pecci--Quinn axion,
rotation and magnetic potential variables. The second formalism is defined
by the same scalar matrix
$P$ and the new vector matrix $\vec \Omega$, both depend only
on the original nondualized string background fields, i.e., on
Calb--Ramond, electromagnetic vector and the metric ones.

It is shown that the supplementary imposition of
the axial symmetry leads to the
vanishing of two components of vector matrix $\vec \Omega$. The remaining third
component $\Omega$ is connected with matrix $Q$ by the system of differential
equations (the dualization relations); their compatibility conditions are
equivalent to the matrix $Q$ motion equation.

Subsequently  the algebraical complex transformation which
directly maps the stationary axisymmetric EMDA equations, written using
the matrices $P$ and $Q$, into the expressed in terms of $P$ and $\Omega$
ones is constructed.
It generalizes the Kramer--Neugebauer transformation \cite {kn1},
of the pure Einstein theory, which was used for the Geroch group \cite {ger}
derivation \cite {kin}.

Also it is established that the matrices $P$ and $\Omega$ define the Gauss
decomposition of the new chiral matrix $N$ which is the $4 \times 4$ EMDA
analogy of the $2 \times 2$ vacuum one. The chiral equation in terms of this
last matrix was the Belinsky and Zakharov \cite {bz1}--\cite {bz2} starting point
for the $2N$--soliton solution construction using inverse scattering
transform technique.

\section{Matrix Formulations of the Stationary String Gravity Equations}

Let us discuss low energy effective four--dimensional action, which
describes the bosonic sector of the heterotic string, taking into account the
gravitational, electromagnetic, dilaton and Calb--Ramond fields:
\begin{eqnarray}\label{e1}
S = \int d^4x {\mid g \mid}^{\frac {1}{2}} (-R+2{\partial \phi}^2
-e^{-2\phi}F^2 +\frac {1}{3}e^{-4\phi }H^2),
\end{eqnarray}
where $R=R^{\mu \nu}_{..\mu \nu}$ is the Ricci scalar
$(R^{\mu}_{.\nu \lambda \sigma} =
\partial _{\lambda}\Gamma ^{\mu}_{\nu \sigma}...)$
of the 4-metric $g_{\mu \nu}$, signature $+ - - -$, $\mu = 0,...,3$ and

\begin{eqnarray}\label{e2}
F_{\mu \nu} &=& \partial _{\mu}A_{\nu}-\partial _{\nu}A_{\mu},
\nonumber \\
H_{\mu \nu \lambda} &=& \partial _{\mu}B_{\nu \lambda}
-A_{\mu}F_{\nu \lambda}
+cyclic.
\end{eqnarray}
Here the scalar field $\phi$ is the dilaton one, and
$B_{\mu \nu}$  we understand as the
antisymmetric tensor Calb--Ramond field.

Further on it will be important
to introduce the pseudoscalar axion field $\kappa$:
\begin{equation}
\partial _{\mu}\kappa = \frac {1}{3}e^{-4\phi}
E_{\mu \nu \lambda \sigma}H^{\nu \lambda \sigma}.
\end{equation}

Below we will study the stationary case when both the metric and
the matter fields are time independent.
As it has been done, \cite{iw}, the four--dimensional line
element can be parametrized according to
\begin{equation}\label{e3}
ds^2=f(dt-\omega _idx^i)^2-f^{-1}h_{ij}dx^idx^j,
\end{equation}
where $i=1,2,3$.
It has been shown before
\cite {jmp}
that in this case part of the Euler--Lagrange equations can be used for the
transition from both spatial components of the vector potential $A_i$
and  functions $\omega _i$ entered
in (\ref{e3}) to the magnetic $u$ and rotation $\chi$
potentials respectively. The new and old variables are connected by
differential relations:
\begin{equation}\label{e4}
\nabla u=fe^{-2\phi}(\sqrt{2}\nabla \times \vec A+\nabla v \times \vec \omega)
+\kappa \nabla v,
\end{equation}
\begin{equation}\label{e5}
\nabla \chi =u\nabla v-v\nabla u -f^2\nabla \times \vec \omega.
\end{equation}
The new notation  $v=\sqrt{2}A_0$ is entered and the three--dimensional operator
$\nabla$ is
corresponded to the three--dimensional metric $h_{ij}$. Also it has been found
that variational equations
for the action (\ref{e1}) are at the same time Euler--Lagrange
equations for the three--dimensional action
\begin{equation}\label{e6}
^3S=\int d^3x h^{\frac {1}{2}}(-^3R+^3L).
\end{equation}
Here $^3R$ is the curvature scalar constructed according to 3--metric
$h_{ij}$ and $^3L$ is
the three--dimensional Lagrangian expressed in terms of $f, \chi, u, v, \phi,
\kappa$ which
is invariant  under the ten--parametric continuous transformation
group isomorphic to $Sp(4,R)$. As it was established \cite {pr2}, $^3L$ can be
written with the aid of the four--dimensional matrix $M$ in the form
\begin{equation}\label{e8}
^3L=\frac {1}{4}Tr(J^M)^2,\qquad J^M=\nabla M M^{-1},
\end{equation}
where $M$, being the matrix of the coset $Sp(4,R)/U(2)$, has the
symplectic and symmetric properties,
\begin{equation}\label{e9}
M^TJM=J,\qquad M^T=M,
\end{equation}
and
\begin{eqnarray}
J=\left (\begin{array}{crc}
0&-I\\
I&0\\
\end{array}\right ).
\end{eqnarray}
Here the matrix $M$ is defined by the Gauss decomposition
\begin{eqnarray}
M=\left (\begin{array}{crc}
P^{-1}&P^{-1}Q\\
QP^{-1}&P+QP^{-1}Q\\
\end{array}\right ),
\end{eqnarray}
where two symmetric matrices $P$ and $Q$ are \cite {pl}
\begin{eqnarray}
P=\left (\begin{array}{crc}
f-v^2e^{-2\phi}&-ve^{-2\phi}\\
-ve^{-2\phi}&-e^{-2\phi}\\
\end{array}\right ),
\end{eqnarray}
\begin{eqnarray}
Q=\left (\begin{array}{crc}
-\chi +vw&w\\
w&-\kappa\\
\end{array}\right ),
\end{eqnarray}
where $w=u-\kappa v$.

It is easy to see that the chiral matrix equation
\begin{equation}
\nabla J^M=0,
\end{equation}
which follows from (8), is equivalent to the system
\begin{eqnarray}
\nabla [P^{-1}(\nabla Q)P^{-1}]=0,\nonumber \\
\nabla [(\nabla P)P^{-1}+QP^{-1}(\nabla Q)P^{-1}]=0.
\end{eqnarray}
After the introduction of two matrix currents
\begin{equation}
J^P=(\nabla P)P^{-1}, \qquad J^Q=(\nabla Q)P^{-1}
\end{equation}
in terms of which equations (15) can be rewritten as
\begin{equation}
\nabla J^Q-J^PJ^Q=0, \qquad \nabla J^P+(J^Q)^2=0,
\end{equation}
for the Einstein equations one has
\begin{equation}
^3R_{ij}=2Tr(J^P_iJ^P_j+J^Q_iJ^Q_j).
\end{equation}
As it can be easily verified, equations (17)--(18) form
the Lagrange system for
the three--dimensional action
\begin{equation}
^3S=\int d^3xh^{\frac {1}{2}}(-^3R+2Tr[(J^P)^2+(J^Q)^2]).
\end{equation}
This makes two symmetric matrices $P$ and $Q$ together with three--metric
$h_{ij}$ the complete set of Lagrange variables for the stationary system
under consideration.

The first relation from (15) can be used for introduction of the
new vector matrix
variable $\vec \Omega$
\begin{equation}
\nabla \times \vec \Omega = P^{-1}(\nabla Q)P^{-1},
\end{equation}
which satisfies the equation
\begin{equation}
\nabla \times [P(\nabla \times \vec \Omega)P]=0,
\end{equation}
as it is immediatelly follows from this definition. Also the second `material'
equation (15) can be written only in terms of the matrices $P$, and
$\vec \Omega$:
\begin{equation}
\nabla [(\nabla P)P^{-1}]+P(\nabla \times \vec \Omega)P\nabla
\times \vec \Omega=0.
\end{equation}

Subsequently, using the differential relations (3), (5) and (6),
one can solve the
equation (20) and express matrix $\vec \Omega$
in terms of the original (nondualized) metric,
electromagnetic and Calb--Ramond variables:
\begin{eqnarray}
\vec \Omega =\left (\begin{array}{crc}
\vec \omega&-(\vec A+A_0\vec \omega)\\
-(\vec A+A_0\vec \omega)&-\vec B+A_0(\vec A+A_0\vec \omega)\\
\end{array}\right ),
\end{eqnarray}
where $B_i=2B_{0i}$ (and all three--dimensional upper indices
are connected with the
metric $h^{ij}$). As it can be
derived from (3), the stationary condition allows
to define the remaining components $B_{ij}$ of the Calb--Ramond field.
Namely, the time independance of $\kappa$ is
equivalent to the relation
\begin{eqnarray}
\nabla \vec C &=& \nabla (\vec B\times \vec \omega)+[\vec B-A_0
(\vec A+A_0\vec \omega)]\nabla \times \vec \omega \nonumber\\
&+&(\vec A+A_0\vec \omega)\nabla \times (\vec A+A_0\vec \omega),
\end{eqnarray}
where $C^i=E^{ijk}B_{jk}$. Thus the spatial components $B_{ij}$
are nondynamical in the stationary case.

It is convinient to introduce the new matrix current
\begin{equation}
J^{\vec \Omega}=P\nabla \times \vec \Omega.
\end{equation}
The Einstein equations (18) in view of (20) obtain the form
\begin{equation}
^3R_{ik}=2Tr(J^P_iJ^P_k+J^{\vec \Omega}_iJ^{\vec \Omega}_k),
\end{equation}
and for written above (21)--(22) we have:
\begin{equation}
\nabla J^P + (J^{\vec \Omega})^2=0, \qquad \nabla \times J^{\vec \Omega} -
J^{\vec \Omega}\times J^P=0.
\end{equation}
The last three equations
also form the complete Lagrange system for the three--dimensional action
\begin{equation}
^3S=\int d^3xh^{\frac {1}{2}}(-^3R+2Tr[(J^P)^2-(J^{\vec \Omega})^2]),
\end{equation}
that means the existense of the alternative non--sigma--model matrix
formulation of the stationary EMDA theory.


\section{Kramer--Neugebauer transformation
in the Stationary Axisymmetric Case}

Now let us turn our attention to the stationary and axisymmetric case, when
the three--dimensional line element can be chosen in the
Luis--Papapetrou form \cite {kin}
\begin{equation}
(dl_3)^2=e^{2\gamma}(d\rho ^2+dz^2)+\rho ^2d\varphi ^2,
\end{equation}
where the function $\gamma$ depends on two space variables
$\rho$ and $z$. Also we
assume that all the remaining field components are independent
of the angular
coordinate $\varphi$.
So the equations for matrices $P$ and $Q$ can be rewritten in the form
\begin{eqnarray}
\nabla [\rho P^{-1}(\nabla Q)P^{-1}]=0,\\
\nabla [\rho ((\nabla P)P^{-1}+QP^{-1}(\nabla Q)P^{-1})]=0.
\end{eqnarray}
Here and further on the operator $\nabla$ is connected with
the two--dimensional
metric $(dl_2)^2=d\rho ^2+dz^2$, and hence it is equivalent to a usual partial
derivative.
These equations are the variational ones for the
two--dimensional action
\begin{equation}
^2S=\int d\rho dz \rho Tr[(J^P)^2+(J^Q)^2].
\end{equation}
The corresponding Einstein equations are transformed into the system
of two relations which define the function $\gamma$:
\begin{eqnarray}
\gamma _{,z} &=& \frac {\rho}{2}Tr[J^P_{\rho}J^P_{z}+J^Q_{\rho}J^Q_{z}],
\nonumber \\
\gamma _{,\rho} &=& \frac {\rho}{4}Tr[(J^P_{\rho})^2-(J^P_{z})^2 +
(J^Q_{\rho})^2-(J^Q_{z})^2],
\end{eqnarray}
where the components of the
matrix currents are defined as before.

As in the stationary case, the equation (30) can be used for
the introduction of
the new symmetric matrix variable $\Omega$,
\begin{equation}
\nabla \Omega=\rho P^{-1}(\tilde \nabla Q)P^{-1},
\end{equation}
where the dual conjugated operator $\tilde \nabla$,
for which $\tilde \nabla _1=\nabla _2$ and $\tilde \nabla _2=-\nabla _1$,
is entered in accordance to [17].
The direct calculation of $\Omega$ leads to its
evident form and also, as it can be seen from
(20),
\begin{equation}
\Omega = (\vec \Omega )_3.
\end{equation}
(Here it was denoted that $x^3=\varphi, \omega _3\equiv \omega$, etc).
The definition (34) forms the compability conditions for the
existense of the matrix $Q$, so that their fulfilment leads to the equation
for matrix $\Omega$:
\begin{equation}
\nabla [\rho ^{-1}P(\nabla \Omega)P]=0.
\end{equation}
The second `material' equation can be also expressed in terms of
$P$ and $\Omega$:
\begin{equation}
\nabla [\rho (\nabla P)P^{-1}+\rho ^{-1}P(\nabla \Omega)P\Omega]=0,
\end{equation}
and it is easy to verify that the system (36)--(37) is the Lagrange one for
the two--dimensional action
\begin{equation}
^2S=\int d\rho dz Tr[\rho (J^P)^2-\rho ^{-1}(J^{\Omega})^2].
\end{equation}
Subsequenty, the differential relation (34) allows to obtain the equations for
the function $\gamma$ in the form
\begin{eqnarray}
\gamma _{,z} &=& \frac {1}{2}Tr[\rho J^P_{\rho}J^P_{z} -
\rho ^{-1}J^{\Omega}_{\rho}J^{\Omega}_{z}],
\nonumber \\
\gamma _{,\rho} &=& \frac {1}{4}Tr[\rho ((J^P_{\rho})^2-(J^P_{z})^2) +
\rho ^{-1}((J^{\Omega}_{\rho})^2-(J^{\Omega}_{z})^2)].
\end{eqnarray}
And so the EMDA system has two alternative descriptions which are
connected with the using of the matrices $Q$ and $\Omega$
in the stationary and axisymmetric case.

The sets of two formulations (32)--(33) and (38)--(39) have the evident
formal analogy with the corresponding systems of equations (with dualized and
nondualized variables) of the pure Einstein theory. Namely, the
Einstein--Maxwell--dilaton--axion theory can be obtained from Einstein one
in the stationary axisymmetric case with the aid of replacements
\begin{eqnarray}
f\longrightarrow P, \nonumber \\
\chi \longrightarrow Q,
\end{eqnarray}
and
\begin{eqnarray}
\omega \longrightarrow \Omega,
\end{eqnarray}
(and the generalization for the stationary case is connected
only with the change
of (41) to $\vec \omega \longrightarrow \vec \Omega$).

As it has been pointed out by Kramer and Neugebauer for the Einstein
system \cite {kn1},
the dualized and nondualized
equation systems can be $directly$ transformed one into another
with the help of the discrete (complex) transformation
\begin{eqnarray}
f &\longrightarrow & \rho f^{-1}, \nonumber \\
\chi & \longrightarrow & i\omega.
\end{eqnarray}
One can check that the formal analogy expressed by the
formulae (40)--(41) allows to
establish the corresponding transformation for the EMDA system. Namely, the
transformation
\begin{eqnarray}
P\longrightarrow \rho P^{-1}, \nonumber \\
Q\longrightarrow i\Omega
\end{eqnarray}
directly maps the pair of equations (30)--(31) into the pair (36)--(37).
By natural reasons we will name it as the Kramer--Neugebauer
transfortmation.

For the complete description it is important to know
the relation,
which reflects the function $\gamma$ behaviour under this transformation.
Let us denote
this function as $\gamma ^Q$ when it is connected with $Q$ and as
$\gamma ^{\Omega}$ when it is defined by $\Omega$.
As it can be established by the straightforward calqulation, the function
$\gamma ^\Omega$ is connected to the function $\gamma ^Q$ as follows:
\begin{equation}
e^{2\gamma ^{\Omega}}=\frac {\rho}{\mid det P\mid}e^{2\gamma ^Q}.
\end{equation}
The difference of this formulae from the Kramer--Neugebauer ones
is defined by the
$2\times 2$ matrix dimension of the EMDA dynamical variables.

As it has been written before, the equations of motion allow
the Sp(4,R)/U(2) matrix representation. In
two--dimensional case these equations have the form
\begin{eqnarray}
\nabla [\rho J^M]=0;
\end{eqnarray}
and
\begin{eqnarray}
\gamma _{,z} &=& \frac {\rho}{4}Tr[J^M_{\rho}J^M_z],\nonumber \\
\gamma _{,\rho} &=& \frac {\rho}{8}Tr[(J^M_{\rho})^2-(J^M_z)^2].
\end{eqnarray}
The same system for Einstein theory is connected to SL(2,R)/SO(2) coset
representation and well known in the literature.

One can try to unite $2\times 2$ matrices $P$ and $\Omega$ into the single
$4\times 4$ one, as it has been done for $P$ and $Q$ with the help of the Gauss
decomposition. The same procedure is possible in the
case of the pure Einstein system (stationary and axisymmetric).
The established analogy (40)--(41) immediately allows
to find the such non--sigma--model matrix representation.
Actually, let us
define the real symmetric matrix $N$ as follows:
\begin{eqnarray}
N =\left (\begin{array}{lcr}
-P&P\Omega\\
\Omega P&\rho ^2P^{-1}-\Omega P\Omega\\
\end{array}\right ).
\end{eqnarray}
It is easy to verify that this matrix satisfies the nongroup condition
\begin{equation}
NJN=-\rho ^2J
\end{equation}
and the chiral equation for it
\begin{equation}
\nabla [\rho J^N]=0
\end{equation}
is equivalent to the pair of equations (36)--(37). (Here the matrix current
$J^N=(\nabla N)N^{-1}$ has been entered). The relation (48) is the natural
generalization of the equality $det N=-\rho ^2$ for the vacuum case.

The equation like (49) was used by Belinsky and Zakharov for the construction
of the $2N$--soliton solution with the help of
the inverse scattering transform technique for the stationary axisymmetric
Einstein equations. As the form of EMDA equations is the same as the Belinsky--
Zakharov ones, the EMDA system allows to make the same procedure as in the
Einstein theory.
The corresponding results will be presented in the next articles.

For the description completeness it is necessary to rewrite the equations (39)
defining the metric function
$\gamma$ into the form where only the matrix $N$ is used.
Such form can be obtained
with the aid of (40)--(41) from the Belinsky--Zakharov one and
it is defined by the relations
\begin{eqnarray}
\Gamma _{,z} &=& \frac {\rho}{8}Tr[J^N_{\rho}J^N_{z}],\nonumber \\
\Gamma _{,\rho} &=& \frac {\rho}{4}Tr[(J^N_{\rho})^2-(J^N_{z})^2],
\end{eqnarray}
where
\begin{equation}
\Gamma = \gamma -\frac {1}{2}ln\mid det P\mid +ln\rho.
\end{equation}
The introduced scalar function $\Gamma$ is evidently connected with
Belinsky--Zakharov one $f$ (which is equal to $e^{2\gamma}\rho f^{-1}$ in
our notations).


\section{Conclusion}
In this article we have presented the descrete complex transformation,
which generalizes the well known Kramer--Neugebauer one of the pure
Einstein theory to the case of the EMDA system. This transformation
directly maps the (dualized) $Sp(4,R)/U(2)$ coset representation of the
stationary axisymmetric EMDA equations into the form, based on the use
of the original (nondualized) string background field components. As it
is established, the nondualized representation also admits the chiral
$4 \times 4$ matrix form which generalizes the formulation
used by Belinsky and
Zakharov for the $2N$--soliton solution construction in the vacuum case.
The same procedure for the system under consideration will be presented in
forthcoming publications.


\acknowledgments

This work was supported in part by the ISF Grant No. M79000.


\end{document}